\documentclass[preprint2]{emulateapj}
\usepackage{graphicx}
\usepackage{natbib}
\def\la{\mathrel{\mathpalette\fun <}}

\def\fun#1#2{\lower3.6pt\vbox{\baselineskip0pt\lineskip.9pt
  \ialign{$\mathsurround=0pt#1\hfil##\hfil$\crcr#2\crcr\sim\crcr}}}

\shorttitle{Signatures of inductive acceleration in AGN jets}
\shortauthors{Ouyed, Sigl, Lyutikov}

\begin{document}

\title{Ultra High Energy Cosmic ray and $\gamma$-ray signatures\\ of 
inductive acceleration in AGN jets}

\author{Rachid Ouyed\altaffilmark{1}, G\"unter Sigl\altaffilmark{2,3}, Maxim Lyutikov\altaffilmark{4}}

 \altaffiltext{1}{Department of Physics and Astronomy, University of Calgary, Calgary,
 Alberta  T2N 1N4, Canada} 
 
  \altaffiltext{2}{APC (AstroParticules et Cosmologie,
  UMR 7164 [CNRS, Universit\'e Paris 7, CEA, Observatoire de Paris]),
10, rue Alice Domon et L\'eonie Duquet, 75205 Paris Cedex 13, France} 
  
  \altaffiltext{3}{Institut d'Astrophysique de Paris, UMR 7095 CNRS -
  Universite Pierre \& Marie Curie, 98 bis boulevard Arago, F-75014 Paris, France}
 
  \altaffiltext{4}{Department of Physics, Purdue University, 525 Northwestern Avenue
West Lafayette, IN, USA} 

\email{ouyed@phas.ucalgary.ca}

\begin{abstract}
The highest energy cosmic rays could be produced by drifts in magnetized, cylindrically collimated, sheared jets of powerful active galaxies (i.e. FR II radiogalaxies; radio loud quasars and high power BL Lacs). We
show that in such scenarios proton synchrotron radiation can give rise
to detectable photon fluxes at energies ranging  from hundreds of keV 
to tens of  MeV.
\end{abstract}

\section{Introduction}

Proton-synchrotron emission has been considered in the past in a context
 involving X-ray flares in Markarian and in
  attempts at explaining the diffuse X-ray background (Muecke\&Protheroe 1999; Aharonian 2002).
   Here we investigate the synchrotron emission from UHE protons
   accelerated in AGN jets by a novel mechanism involving inductive
    fields (Lyutikov\&Ouyed 2005, hereafter LO5). This is a scheme of Ultra-High Energy Cosmic Ray (UHECR) acceleration due to drifts in magnetized, cylindrically collimated, sheared jets of powerful active galaxies (with jet power in Poynting flux $\geq 10^{46}$ erg s$^{-1}$) - It is important to note that the acceleration mechanism suggested in LO5 is associated with the jets of AGNs, and not
    with their hot spots. We pointed out 
    in LO5  that a positively charged particle carried by such a plasma is in an unstable equilibrium if ${\bf B} \cdot \nabla \times {\bf v}< 0$, so that kinetic drift along the velocity shear would lead to fast, {\it  regular} energy gain; here ${\bf B}$ is the jet magnetic field dominated by the toroidal
    component while  ${\bf v}$ is indirectly related to the angular velocity of the black-hole-accretion
    disk system driving the jet and is falling off with the cylindrical radius.
     We showed that if a seed of pre-accelerated particles with energy below the ankle $\leq 10^{18}$ eV is present, these particles can be boosted to energies above $ 10^{19}$ eV. A key feature of the mechanism is that the highest rigidity particles are accelerated most efficiently implying the dominance of light nuclei for extragalactic cosmic rays in our model: from a mixed population of pre-accelerated particle the drift mechanism picks up and boosts protons preferably.

    In this acceleration mechanism synchrotron radiation is thus a natural outcome
    since the protons escape the jet sideways in a direction transverse
    to the magnetic field in the jet. In this letter we propose that the resulting
     radiation spectrum makes an attractive  interpretation
      of non-thermal X-ray and $\gamma$-ray emission from powerful AGNs and might be 
       a direct indication of inductive acceleration in AGN jets
         as a viable mechanism for  UHECR acceleration.

\section{Synchrotron emisison}

At a distance $r$ from the central engine producing X-rays of energy
$\epsilon$ with a luminosity $L_{\rm X}$, the density of these X-ray
photons is
\begin{eqnarray}\label{eq:n_X}
n_{\rm X}&\simeq& \frac{L_{\rm X}}{4\pi r^{2} \epsilon}
\sim 1.7\times 10^{-6} \ {\rm cm}^{-3}\\\nonumber
&\times&\left(\frac{L_{\rm X}}{10^{47}\ {\rm erg s}^{-1}}\right) \left(\frac{\epsilon}{10 \ {\rm keV}}\right)^{-1}\left(\frac{r}{\rm Mpc}\right)^{-2}\ .
\end{eqnarray}

For a proton-$\gamma$ cross-section $\sigma_{\rm N\gamma}\sim 100 \mu b$
the corresponding nucleon mean-free path in this X-ray background is
 \begin{equation}
 l_{\rm N\gamma}\sim 1.9\times 10^{9}\ {\rm Mpc}\ \left(\frac{L_{\rm X}}{10^{47}\ {\rm erg s}^{-1}}\right)^{-1}\left(\frac{\epsilon}{10 \ {\rm keV}}\right)\left(\frac{r}{\rm Mpc}\right)^{2}\ .
 \end{equation}
The cross-section for inverse Compton (IC) scattering is on the other hand
$\sigma_{\rm IC}\sim 2\times 10^{-31}\ {\rm cm}^{2}$, corresponding to
a length scale
 \begin{equation}
 l_{\rm IC}\sim 9.3\times 10^{11}\ {\rm Mpc}\ \left(\frac{L_{\rm X}}{10^{47}\ {\rm erg s}^{-1}}\right)^{-1}\left(\frac{\epsilon}{10 \ {\rm keV}}\right)\left(\frac{r}{\rm Mpc}\right)^{2}\ .
 \end{equation}
 
The optical depth  for $pp$ interaction in the
vicinity of the AGN is much smaller than one.  
This  implies that the neutrino flux is much smaller than the cosmic
ray flux and therefore should not be detected in the forseeable future.
The sensitivity of ICECUBE to the neutrino power of a source at
distance $D$ is $\sim2\times10^{45}\,{\rm erg s}^{-1}(D/100\,{\rm Mpc})^2$
(Ahrens et al. 2004).
 
We are left with synchrotron losses which occur on a length scale of
\begin{equation}
l_{\rm B} \sim 430\ {\rm kpc} \left(\frac{E_{\rm p}}{10^{20}\ {\rm eV}}\right)^{-1}\left(\frac{B}{{\rm mG}}\right)^{-2}\ ,
\end{equation}
where we take qualitatively the jet magnetic field to be of the order of $\sim 100\,\mu$G--$10\,$mG (e.g. Sikora et al. 1997; Aharonian 2001;
Stawarz et al. 2005; Schwartz et al. 2006).
  
The photon energy fluence peaks at the energy (Lang 1980)
\begin{equation}
\label{eq:egamma}
E_{\rm sync} \sim 25\ {\rm MeV} \left(\frac{E_{\rm p}}{10^{20}\ {\rm eV}}\right)^{2}\left(\frac{B}{{\rm mG}}\right)\ ,
\end{equation}
and since in the model of L05,
\begin{equation}
  E_{\rm p}\la4\times10^{20}\,
  \left(\frac{L_{\rm P}}{10^{46}\,{\rm erg}{\rm s}^{-1}}\right)^{1/2}
  \,{\rm eV}\,,
\end{equation}
where $L_{\rm P}$ is the Poynting flux luminosity of the AGN, this implies
\begin{equation}
\label{eq:egamma2}
E_{\rm sync} \la  400\ {\rm MeV} \left(\frac{L_{\rm P}}{10^{46}\,{\rm erg\ s}^{-1}}\right) \left(\frac{B}{{\rm mG}}\right)\ .
\end{equation}
It is important to note that in the model of LO5 the accelerating potential
 is defined by the Poynting flux luminosity, which in some sources (with high
 dissipation rates) might  exceed the {\it observed} bolometric luminosity; intrinsically of
 course the Poynting flux is always smaller than the total luminosity.

Let us define $f_{\gamma}$ as the fraction of the luminosity $L_{\rm UHECR}$
of accelerated UHECR that 
goes into $\gamma$-rays around the energy given by eq.~(\ref{eq:egamma}).
If $l_{\rm ext}$ is the lateral extent of the jet over which the protons are
accelerated, one has $f_\gamma\sim l_{\rm ext}/l_{\rm B}$ for
$l_{\rm ext}/l_{\rm B}\ll1$. Given that typical  jet width is of the order
of $0.1$-$1$ kpc (Hughes 1991) it implies a range of.
\begin{equation}\label{eq:f_gamma}
  f_\gamma\sim10^{-2}\left(\frac{l_{\rm ext}}{\rm 1\ kpc}\right)
  \left(\frac{L_{\rm P}}{10^{46}\,{\rm erg}\,{\rm s}^{-1}}\right)^{1/2}
  \left(\frac{B}{\rm mG}\right)^2\,.
\end{equation}
  
These photons can leave the jet with
negligible interactions: First, photons of energy given by eq.~(\ref{eq:egamma})
are below the threshold for pair production with $\sim10\,$keV X-ray photons.
Second, with $l_{\rm j}$ the jet length, the density of synchrotron $\gamma$-rays is
\begin{eqnarray}
n_\gamma&\sim&\frac{f_\gamma L_{\rm UHECR}}{2\pi E_{\rm synch}l_{\rm j}l_{\rm ext}}
\sim3.5\times10^{-6}\,{\rm cm}^{-3}\\\nonumber
&\times&\left(\frac{f_\gamma L_{\rm UHECR}}{10^{45}\ {\rm erg s}^{-1}}\right)
\left(\frac{\rm 1~MeV}{E_{\rm synch}}\right)
\left(\frac{100~{\rm kpc}}{l_{\rm j}}\right)
\left(\frac{1~{\rm kpc}}{l_{\rm ext}}\right)\,,
\end{eqnarray}
which is comparable or smaller than Eq.~(\ref{eq:n_X}) and,
therefore, does not modify the proton interactions. With
$\sigma_{\rm T}\simeq6.7\times10^{-25}\,{\rm cm}^2$
the Thomson cross section, the optical depth
for pair production among these photons is thus $\tau_{\gamma\gamma}\sim
\sigma_{\rm T}n_\gamma l_{\rm ext}\ll1$.

Below the energy given by eq.~(\ref{eq:egamma}),
the synchrotron spectrum of a single proton scales as
$E^2j_{\rm synch}(E)\propto E^{4/3}$, whereas it cuts off
exponentially above $E_{\rm synch}$.  The total synchrotron
power of a proton of energy $E_{\rm p}$ scales as $E_{\rm p}^2$.
Folding with a charged primary
spectrum $j_p(E_{\rm p})\propto E_{\rm p}^{-\gamma}$, this yields a
 synchrotron flux
$E^2j_{\rm synch}(E)\propto E^{-(\gamma-3)/2}$ up to $E_{\rm synch}$ for
the highest energy protons. In the model of LO5, in the extreme case of an infinite potential, 
$\gamma\simeq2$ and thus $E^2j_{\rm synch}(E)\propto E^{1/2}$ or,
a spectrum $j_{\rm synch}(E)\propto E^{-1.5}$.  Depending on
composition and whether the ankle in the cosmic ray spectrum
is due to a cross-over from a galactic to an extragalactic population,
the source injection spectral index is $2.3\lesssim\gamma\lesssim2.7$
(see, e.g., Berezinsky et al. 2006; Allard et al. 2007). In our model softer injection spectra with $\gamma > 2$ are more realistic since the potential might only be used partially during the inductive acceleration process.
 Note that the synchrotron spectrum is not sensitive to the low
energy continuation of the primary spectrum as long as $\gamma\le3$.

\section{$\gamma$-ray signatures}

For a point source power $L_{\rm UHECR}$ at distance $D$
this results in a $\gamma$-ray point flux of
\begin{eqnarray}
\label{eq:esource}
E_{\rm sync}^{2} j_{\rm s}(E_{\rm sync}) &\simeq& 8\times 10^{-10} \ {\rm erg\ cm}^{-2}\ {\rm s}^{-1}\\\nonumber
&\times& \ f_{\gamma}\times \left( \frac{L_{\rm UHECR}}{10^{45}\ {\rm erg\ s}^{-1}}\right) \left(\frac{D}{100 \rm Mpc}\right)^{-2}\ ,
\end{eqnarray}
where $10^{45}$ erg s$^{-1}$ is about $\sim 1\%$ of the bolometric luminosity of
the brightest AGNs and about 10\% of the jet power
$L_{\rm P}\gtrsim10^{46}(E_{\rm max}/4\times10^{20}\,{\rm eV})^2
\,{\rm erg}\,{\rm s}^{-1}$ required to accelerate protons up to $E_{\rm max}$
of a few 100 EeV in the scenario of LO5. 
Given that $f_{\gamma}\sim 10^{-4}$--$10^{-1}$, this implies that the point and diffuse fluxes are within the sensitivities of existing and upcoming instruments such as
COMPTEL (0.75-30 MeV), EGRET (30 MeV - 20 GeV), SIGMA, INTEGRAL and GLAST
GRB monitor (10 keV - 25 MeV). Note that Eq.~(\ref{eq:esource}) is
just the angle-averaged flux;  since for simplicity we assume that the synchrotron radiation is strongly
beamed perpendicular to the jet, in which case the actual flux will strongly depend on
jet orientation. In reality however the direction of the synchrotron emission depends
 on the structure of the magnetic field and its degree of entanglement as well as 
  aberration effects.

    As discussed in LO5, requirements on luminosity (and the
assumption of a high proton content in UHECRs) excludes low power nearby
sources, like the nearest
AGN M87 and the starburst galaxy M82. Instead high power radio galaxies with
$L\gtrsim10^{45}\,{\rm erg}\,{\rm s}^{-1}$
(e.g. FR II radiogalaxies; radio loud quasars and high power BL Lacs) at intermediate
distances are favored, such as Pictor A ($z=0.035$), PKS 1333-33 ($z=0.0124$), PKS
   2152-69 ($z=0.027$), PKS 1343 ($z=0.012$), and the Seyfert galaxy 3C 120 ($z=0.033$).
   It is still worth considering high power FRI as possible candidates in case
    as we have said their Poynting power turns out to exceed the {\it observed} bolometric value.
 For example, Centaurus A, the nearest
giant radio galaxy at a distance of $\sim3.6\,$Mpc with a jet pointing with $\sim70^\circ$ to the line of sight has been seen by SIGMA and COMPTEL.
 It has a $\gamma-$ray spectrum somewhat harder than $E^{-2}$, roughly
consistent with the $E^{-(\gamma+1)/2}$ $\gamma-$ray spectrum with
$2\lesssim\gamma\lesssim3$, suggested by the 
mechanism discussed here, cutting off above a few MeV. The (time variable)
flux around 100 keV is $\sim2\times10^{-10}\,{\rm erg\ cm^{-2}\ s^{-1}}$
(Bond et al. 1996, Sreekumar 2000). From Eq.~(\ref{eq:esource}), this flux 
density would imply $f_\gamma(L_{\rm UHECR}/10^{43}\,{\rm erg\ 
s^{-1}})\gtrsim2.8$\%. Thus for this source the fraction of $L_{\rm UHECR}$ luminosity that goes into $\gamma$-rays is a few percents, larger than the
value derived from equation (\ref{eq:f_gamma}): Indeed, the total high
energy (we take as Poynting) luminosity of Cen A is estimated to $\sim10^{43}\,{\rm erg\ s^{-1}}$
(Bond et al. 1996, Israel 1998), and thus $f_\gamma\lesssim3\times10^{-4}
(l_{\rm ext}/{\rm kpc})(B/{\rm mG})^2$.
This discrepancy  is likely less stringent for
emission beamed toward the observer which may be the case since the jet
is seen almost sideways; a larger $l_{\rm ext} B^2$ could also
remedy to this discrepancy. The maximal UHECR energy is thus
$E_{\rm p}\sim10^{19}\,$eV. This implies $E_{\rm synch}\sim200\,$keV
according to Eq.~(\ref{eq:egamma}) if $L_{\rm UHECR}$ is comparable,
which could thus explain the $\sim100\,$keV flux with $f_\gamma$
of a few percent.

   Blazars  are of particular interest to our model since they are known to show a spectral bending
    (i.e. spectral index break) around a few MeV to tens of MeV energies. 
    Combination of OSSE, COMPTEL and EGRET
     data  show a spectral bending  during the low
   luminosity  and high-luminosity $\gamma$-ray state of some sources with a
   a break energy  estimated to range between
    $\sim$ 1 MeV and $\sim$ 20 MeV (e.g. Kurfess 1994; Collmar et al. 1997).
    This break energy   lies within the range predicted
    in our model [see eqs (\ref{eq:egamma}) and (\ref{eq:egamma2})]. Furthermore, 
 in time-averaged analyses the spectra of these AGNs are well described by power-law
 shapes ($E^{-\alpha}$) with a photon index $\alpha$ of the order of 2 
   which  can be accounted for in our model. 
   We note however that blazars are at high enough redshift
   to lie  beyond the GZK cut-off.  Thus, these $\gamma$-ray signatures (which
    we argue might be indicative of the inductive acceleration mechanism) will
    have associated UHECRs at most below $\sim5\times10^{19}\,$eV.

If powerful AGNs collectively explain the observed UHECR flux then their 
volume emissivity above $\simeq10^{19}\,$eV is roughly of the order of
$Q_{\rm UHECR}\sim 10^{37}$ erg Mpc$^{-3}$ ${\rm s}^{-1}$, and one obtains
a diffuse gamma ray flux of
\begin{eqnarray}
E_{\rm sync}^2 j_{\rm d}(E_{\rm sync}) &\simeq& f_{\gamma}\times 3\times 10^{-9}
\ {\rm erg\ cm^{-2}\ s^{-1}\ sr^{-1}}\nonumber\\
&\times& \left( \frac{Q_{\rm UHECR}}{10^{37}\ {\rm erg\ Mpc^{-3}\ s}^{-1}}\right)\,.
\label{eq:evolume}
\end{eqnarray}
With $f_\gamma\sim1\,$\%, the  diffuse flux predicted from our scenario could thus also contribute a fraction of a few percent to the diffuse flux
observed by COMPTEL, $\sim3\times10^{-9}\,{\rm erg\ cm^{-2}\ s^{-1}\ sr^{-1}}$
(Sreekumar 2000).

\end{document}